\documentstyle [prl,aps,epsf]{revtex}
\begin{document}
\twocolumn[{
\widetext
\draft

\title{ Exact results for quantum 
phase transitions in random XY spin chains
}

\author{Ross H. McKenzie\cite{email}}

\address{School of Physics, University of New
South Wales, Sydney 2052, Australia}

\date{Received 18 September 1996}
\maketitle
\mediumtext
\begin{abstract}
The effect of disorder on the quantum phase transitions
induced by a transverse field,
anisotropy, and dimerization in XY spin chains is investigated.
The low-energy behavior near
the critical point is described by a
Dirac-type equation for which an exact analytic
treatment is possible.
Results obtained for the dynamical critical exponent,
the specific heat, and transverse susceptibility
agree with results recently obtained using
a real space renormalization group decimation technique.
A non-zero transverse field
changes the universality class of the anisotropy transition.
\\
\\
Submitted to Physical Review Letters.
\\
\end{abstract}

}] \narrowtext


Theoretical studies of quantum phase transitions in the presence of
quenched (i.e., time-independent) disorder has been stimulated by recent
experiments on spin glasses, $^4$He absorbed in porous media,
 and superconductor-insulator transitions in dirty thin films.
  Compared to thermal
phase transitions in disorder-free systems these transitions are poorly
understood because many of the theoretical methods (e.g., exact solutions,
renormalisation group, $\epsilon$ expansions) that have been so useful for
pure systems are difficult to 
implement for disordered systems.  Yet 
these phase transitions are associated with particularly rich physics
such as large differences between average
and typical behavior, new universality classes,
 logarithmic scaling,
Griffiths phases\cite{griffiths} (in which
 susceptibilities diverge  although 
  there are only short-range correlations),
and the breakdown of folklore         
such as ``the correlation length is inversely 
proportional to the energy gap''\cite{ry}.
This Letter considers a simple 
exactly-soluble model which has many of these 
interesting properties.

Fisher recently performed an exhaustive study of the 
effect of randomness on the simplest
spin model
to undergo a quantum phase transition:
the transverse field Ising spin chain\cite{dsf}.  He used
a real space renormalization group decimation technique
(RSRGDT) \cite{ma}
 which he claims is exact near the critical point.
Fisher  found the phase diagram
(which included Griffiths phases near the critical point), all the
critical exponents, and scaling forms for the 
magnetization and correlation functions in an external field.
The latter have never been derived for the disorder-free case
but can be derived in the presence of disorder because 
distributions become extremely broad near the critical point.
The same model was recently studied numerically by
 Young and Rieger\cite{yr}, who
found results consistent with Fisher.
The RSRGDT         has also be used to study
the effect of disorder on dimerized\cite{hyman} and
anisotropic\cite{dsf2} spin chains,
chains with random spin sizes\cite{west},
and quantum Potts and clock chains\cite{senthil}.
 Possible 
experimental realizations of random spin chains
are quinolonium(TCNQ)$_2$\cite{bul}
and Sr$_3$CuPt$_{1-x}$Ir$_x$O$_6$\cite{nguyen}.

This Letter gives
an exact treatment of the quantum phase transitions
in XY spin chains in the
presence of disorder.  The spin chains are mapped onto a fermion
model, the continuum limit of which
is a Dirac-type equation with random mass, for which 
{\it exact} analytic
results can be derived.  The results agree with those
of the RSRGDT\cite{dsf,hyman,dsf2}, supporting 
Fisher's claim      that it is exact
near the critical point\cite{however}.
New results are obtained for the anisotropy 
 transition in a non-zero average transverse field.
It is in a different universality class
to the Ising transition.

The Hamiltonian to be considered is:
\begin{equation}
H = -\sum_{i=1}^L \left(J_i^x \sigma^x_i \sigma^x_{i+1} +
 J_i^y \sigma^y_i \sigma^y_{i+1} +
 h_i \sigma^z_i \right) \ .
\label{ham}
\end{equation}
The $\{\sigma^\alpha_i\}$ are Pauli spin matrices, and the
interactions
 $J_i^x$,  $J_i^y$, 
and transverse fields $h_i$ may be independent
random variables (see Table I)
with a Gaussian distribution.
The  average values will be denoted
\begin{equation}
\langle J_i^x \rangle \equiv J^x
\ \ \ 
\langle J_i^y \rangle \equiv J^y
\ \ \ \ \ \ \
\langle h_i \rangle \equiv h 
\label{mean }
\end{equation}
and will all be assumed to be positive.

At zero temperature and in
the  absence of disorder the model undergoes two
distinct quantum phase transitions \cite{dennijs}.
Both transitions are second order.
The transition  at
$J^x + J^y= h $ from a paramagnetic to an ferromagnetic phase
will be referred to as the {\it Ising transition}\cite{pfeuty}.
The  transition at
$ J^x=J^y $ for $ h < (J^x + J^y) $ from an
Ising ferromagnet with magnetization in the $x$ direction to one with
magnetization in the $y$ direction will be
referred to as the {\it anisotropy transition}\cite{dsf2,lsm}.
The {\it dimerization transition} 
involves the Hamiltonian (\ref{ham})
 with $\langle J_i^x \rangle= \langle J_i^y \rangle =J + (-1)^i\Delta  $
and $\langle h_i \rangle=0$.
At $\Delta =0$ there is a transition
from a spin liquid to a gapped phase with
long-range topological order \cite{hyman}.

Under a Jordan-Wigner transformation which maps spins
 onto spinless fermions
(\ref{ham}) becomes\cite{lsm}
\begin{eqnarray}
 H = &-&\sum_{i=1}^L
\Bigl(  (J_i^x + J_i^y) (c^\dagger_i  c_{i+1} +  c^\dagger_{i+1} c_i)
\nonumber \\
&+&  (J_i^x - J_i^y)
 (c^\dagger_i  c^\dagger_{i+1} - c_i c_{i+1}) + h_i (2c_i^\dagger c_i - 1)
  \Bigr) \label{jw}
\end{eqnarray}
where boundary terms have been neglected
and $c^\dagger_i$ and $c_i$ denote creation
and annihilation operators
for a fermion on site $i$.
 In the disorder-free case
(\ref{jw}) can be diagonalised by a 
Bogoliubov transformation\cite{dennijs,lsm}.

The continuum
limit of (\ref{jw}) must be taken relative to
the Fermi wavevector $k_F$, the wavevector
at which the energy gap in the fermionic spectrum occurs for
the disorder-free case\cite{dennijs}.
  This wavevector depends on which transition
is being considered (see Table I).
The low-energy properties of 
all the transitions in Table I are
described by  \cite{dirac} 
\begin{equation}
H = \int dx \Psi^\dagger  \bigg[-iv_F \sigma^3
{\partial \over \partial x} 
 + V(x) \sigma_+ + V(x)^* \sigma_- 
\bigg] \Psi
\label{hamel}
\end{equation}
where $\sigma^a$ ($a=1,2,3$) and $ \sigma_\pm \equiv {1 \over 2}
(\sigma^1 \pm \sigma^2) $ are Pauli matrices.
$\Psi(x)$ is a spinor, with
(modulo a transformation in
spinor space) the components 
given in Table I.
 $V(x)$ is Gaussian white-noise potential with
\begin{equation}
\langle V(x) \rangle = \Delta 
\ \ \ \ \ \ \
\langle V(x) V(x^\prime)^* \rangle = \gamma \delta (x -x^\prime).
\label{corr}
\end{equation}
$\Delta $ measures the deviation from the  critical point
and $|\Delta|$ is the energy gap in the absence
of disorder.
Table I lists values of $\Delta$, the
Fermi velocity $v_F$, and the random
variables whose standard deviation
equals $\gamma$,
for the Ising, anisotropy, and dimerization  transitions.
For the Ising transition ($k_F=0$) and
the anisotropy and dimerization transitions in
zero transverse field ($k_F=\pi/2$)
$k_F$ is {\it commensurate} with the lattice
and $V(x)$ is real.
It will be seen that the effect of disorder is
significantly different for commensurate and
incommensurate cases.
The continuum limit describes the low-energy properties ($E \ll v_F$)
when $ v_F \gg \Delta, \sqrt{\gamma},$ 
i.e., arbitrarily close to the critical point and for weak disorder.

It is useful to define  an energy $D$
and a dimensionless parameter $\delta$
which are measures of the disorder strength
and the deviation from criticality, respectively 
\begin{equation}
D \equiv {\gamma \over v_F} \ \ \ \ \ \
\delta \equiv {|\Delta| \over D}.       
\label{delta}
\end{equation}
 Note that for the Ising transition
with $J^y=0$, to leading order
in $\Delta/J^x $,
the parameter $\delta$
defined by Fisher\cite{dsf} and Young and Rieger\cite{yr}
is $\Delta/D$.

The advantage of casting the problem in the form of the Hamiltonian
 (\ref{hamel})
is that the latter
has been studied extensively previously, and
{\it exact} analytic expressions given for the energy dependence of the
disorder-averaged 
density of states $ \langle\rho (E)\rangle$ and the localization length
$\lambda(E)$\cite{ov,halp,gor,golub,hayn,john,mertsching,mck,bouch}.
(Due to the one-dimensionality all the states are localized
by the disorder).
 The exact results have been found by   
Fokker-Planck equations\cite{ov,halp}, supersymmetry\cite{hayn,john},
 the replica trick\cite{bouch},
S-matrix summation \cite{golub}, and the Dyson-Schmidt method
\cite{mertsching}.
 The localization length can be found because in one dimension
it is related to the real part of the one-fermion  
 Green's function\cite{john,thou}.
The density of states  and the localization length
can be written in terms of $f_{\delta} ^\prime (u),$
the derivative of a dimensionless function $f_{\delta}(u)$
given in Table  II,
\begin{equation}
 {d \over dE} {1 \over \lambda(E)} + i \pi \rho (E)
        = \pi \rho_0 f_{\delta} ^\prime ( E / D) 
\label{gf}
\end{equation}
where
$\rho_0 \equiv 1/(\pi v_F)$ is the value of the density of
states at  high energies ($|E| \gg \Delta, D )$.

The low-energy   
 ($|E|  \ll D          $)
behavior of the density of states is given
in Table II.
For the commensurate case the
density of states diverges at $E=0$ for  $\delta < 1/2$
  and is zero at $E=0$ for $\delta > 1/2 $.
This difference
 will lead to qualitatively very different behaviour for these two
cases. In the former case some susceptibilities will
diverge as the temperature approaches zero. This
corresponds to a Griffiths phase \cite{griffiths}.
 The corresponding phase diagram for the
Ising transition is shown
in Fig.  \ref{figphased2}.
The full energy dependence of the density of states for
various values of  $\delta$ is shown in
Fig. \ref{figdos}.


Distinctly different 
behavior  occurs for the incommensurate case.
The density of states is always finite at
zero energy.
There is a smooth crossover from gapless behavior ($\langle \rho(0)
\rangle \sim \rho_0$
 for $\delta < 1$) to effectively gapped behavior ($\langle \rho(0)
\rangle \ll \rho_0$ for $\delta \gg 1$). 
Hence the Griffiths phases still exist
but no longer have  clearly defined boundaries.
Similarily, for the commensurate case
more than one type of disorder (e.g., the anisotropy
transition with     both a
random transverse field and  random anisotropy)
removes the singularity in the density of states\cite{ov}.

{\it The specific heat.}
Because the eigenstates of the Hamiltonian 
(\ref{hamel}) are non-interacting fermions
the low-temperature behavior of the specific
heat and the transverse susceptibility
(for the dimerization and anisotropy transitions)
 follows from the energy
dependence of the disorder-averaged 
density of states\cite{bul,smith}
and are given in Table II.
The transverse susceptibility clearly shows
a Griffiths singularity.
The results for the commensurate case agree with
results obtained using the RSRGDT\cite{dsf,hyman,dsf2}, 
supporting Fisher's claim \cite{dsf} that it             
is exact near the critical point.

{\it Dynamical critical exponent $z$.}
This relates the scaling of energy (or time) scales
to length scales.
A crude scaling argument\cite{scale} implies that
$   \langle \rho(E) \rangle  \sim E^{1/z -1}.$
Thus for the commensurate case, to leading order in $\delta$,
$z= 1 / (2 \delta )$,
in agreement with Fisher\cite{dsf} and Young and Rieger\cite{yr}.
This is a particularly striking result because it shows
that (i)    $z$ is not universal and
(ii)    $z$     diverges at the critical point.
The latter implies logarithmic scaling and
activated dynamics\cite{ry}.
In contrast for the incommensurate case,
the density of states is finite and constant at low energies
and so $z=1$, as in the absence of disorder.

{\it Finite size scaling.}
Monthus et al.\cite{mon} studied          
an equation equivalent to (\ref{hamel})
with $V(x)$ real and $\Delta=0$ \cite{bouch}.
They have shown that  on a line of length $L$,
for a typical potential $V(x)$
the lowest eigenvalue $E_0$ scales like
$E_0^2 \sim \exp(-L^{1/2}).$
This is consistent with the scaling of
$\ln E_0$ with $L^{1/2}$ at the critical point
found numerically by Young and Rieger\cite{yr}.
The average $\langle E_0^2 \rangle \sim \exp(-L^{1/3})$ \cite{mon},
showing the discrepancy between {\it average} and
{\it typical} values.

{\it Correlation lengths.} Fisher\cite{dsf} stressed the
distinction between average and
 typical correlations.
If $C_{ij}\equiv \langle A_i   A_j \rangle$
denotes a correlation function of a variable $A_i$
then the average correlation function
$C_{av}(r) \equiv {1 \over L} \sum_{i=1}^L C_{i,i+r}$
is what is measured experimentally.
Away from the critical point $C_{av}(r) \sim \exp(-r/\xi_{av})$
where $\xi_{av}$ is the average
correlation length.
However, $C_{av}(r)$ is dominated by rare 
pairs of spins with $C_{ij} \sim 1.$
In contrast, with probability one $C_{i,i+r}
 \sim \exp(-r/\xi_{typ})$
where $\xi_{typ}$ denotes the typical
correlation length. It  is distinctly different
from $\xi_{av}$ ($\xi_{typ} \ll \xi_{av} $),
 having  a different critical exponent.
The localization length is useful because it is proportional
to the typical correlation
length for quantitites that are
diagonal in the fermion representation\cite{klein}.
  Consequently the 
results in Table II imply that for the
commensurate case the
typical correlation length,
$\xi_{typ} \sim \lambda(0)^{-1} \sim \Delta^{-1}$,
consistent with previous work \cite{dsf,sm}. This critical exponent is
not modified by the presence of disorder.
In contrast, for the incommensurate case,
$\lambda(0)$ is {\it finite } at the critical point.

{\it Distribution functions.}
Both Fisher\cite{dsf} and Young and Rieger\cite{yr}
considered the distribution functions for
various quantities. 
The present approach can be used to obtain exact results
by using known results 
for the distribution
function for the density of states of
one-dimensional conductors\cite{altshuler}.
The cumulants (or irreducible moments)
of the transverse susceptibility 
for the incommensurate case  at criticality are
\begin{equation}
\langle (\chi_{zz})^n\rangle_c 
= \chi_0^n {\Gamma({3 \over 2}) \Gamma(n)^2
\over  \Gamma(n+{1 \over 2})}
 \left( {\pi D      \over 16 T } \right)^{n-1}
\label{mom}
\end{equation}
where $\chi_0$ is the susceptibility in the
absence of disorder.
These moments completely determine the
distribution function which has been
extensively analyzed in Ref. \onlinecite{altshuler}.
For $T \gg D         $ the distribution is Gaussian
and centered on the mean.
In contrast, for 
$T \ll D $ the distribution is extremely
broad and asymmetric.
The maximum of the distribution then occurs near
$ \chi_0 \exp (-1/s)/2s$, where $s=16 T/(D \pi),$
 which is much less than the mean $\chi_0$.
This shows the large discrepancy between 
typical and average values.
Hopefully, (\ref{mom})  can be generalized to the commensurate
case and away from criticality.


This work    was supported by the Australian Research Council.
I thank D.D. Betts, R.J. Bursill,  D.S. Fisher,
M. Gulacsi, G. Honner, and V. Kotov for very helpful discussions.
D. Scarratt produced the figures.


\begin{figure}
\centerline{\epsfxsize=8.0cm \epsfbox{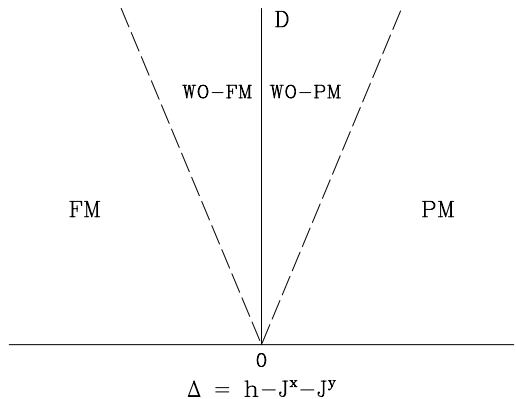}}
\vskip -2.0 truecm
 \caption{
Phase diagram of the Ising transition          
in a random transverse field. The horizontal
axis is a measure of the deviation from criticality in the non-random
model. The vertical axis is the amount of disorder: the rms
fluctuation in the transverse field.  
The four phases are: ferromagnet(FM),
 weakly ordered ferromagnet (WO-FM),
weakly ordered paramagnet (WO-PM),
 and paramagnet (PM).
 The weakly ordered phases are Griffiths phases
\protect\cite{griffiths}.
\label{figphased2}} \end{figure}

\vskip 13cm

\begin{figure}
\centerline{\epsfxsize=8.0cm \epsfbox{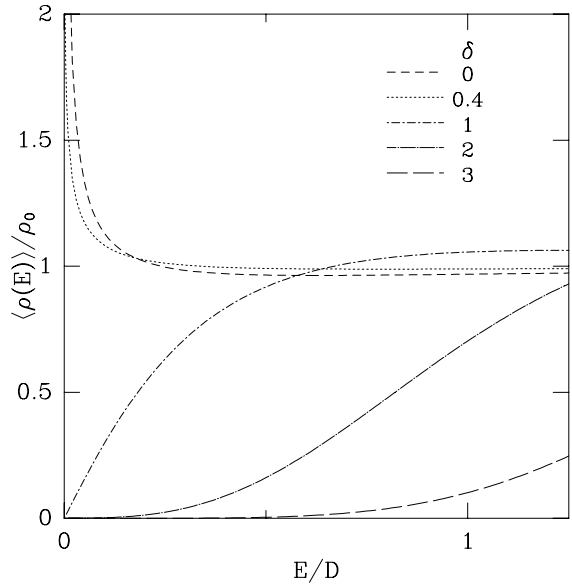}}
\vskip 0.3cm
\caption{
Energy dependence of the disorder-averaged
density of states for the commensurate case
for various values of the dimensionless parameter
$\delta$
 (see Eq. (\protect\ref{delta})),
which is a measure of the deviation from criticality.
 The density of states is singular at the Fermi energy $(E=0)$
 when $\delta < {1 \over 2}.$
This parameter range corresponds to a Griffiths phase.
Note that only far from criticality
 ($\delta \gg 1$) is there effectively a gap in
the system.
  This contrasts with
the disorder-free case, for which
there is always a gap except at the critical point.
\label{figdos}}
\end{figure}

\newpage
\widetext

\begin{table}
\caption{
Different parameters for the continuum limit
of three different transitions. Near the 
critical point all are described by the
 low-energy effective Hamiltonian (\protect\ref{hamel}).
The lattice constant is set to unity.
The Fermi wavevector $k_F$ determines whether
the problem is commensurate or not.
$v_F $ is the Fermi velocity.
$\Delta$ is a measure of the deviation from
criticality.
The standard deviation of the random variable
equals $\gamma$ which enters Eq. (\protect\ref{corr}).
}
\begin{tabular}{lccccc}
Transition & $\cos k_F$ & $v_F$ & $\Delta$ & $\Psi(x=n)^\dagger$
 & Random
variable \\
\tableline
Ising & 1 & $| J^x - J^y |$ & $ h -  J^x - J^y$
 & $(c_n^\dagger, c_n)$ & $h_i$\\
Anisotropy & $ \displaystyle {h \over J^x + J^y}$ & $(J^x + J^y)\sin k_F$
 & $ (J^x - J^y)\sin k_F $
 & $(c_n^\dagger e^{-ink_F}, c_n e^{-ink_F})$ & $J^x_i-J_i^y$\\
Dimerization & 0 & $2J$ & $\Delta$ 
 & $(c_n^\dagger i^{-n}, c_n^\dagger i^n)$ & $J^x_i=J_i^y$\\
\end{tabular}
\label{table1}

\end{table}

\begin{table}
\caption{
Summary of the low-energy behavior of
different physical quantities.
The commensurate case corresponds to the Ising
transition and to the anisotropy and dimerization
transitions in zero transverse field. The incommensurate
case corresponds to the anisotropy 
transition in a non-zero transverse field.
For the commensurate case qualitatively different
behavior occurs at the critical point ($\delta=0$)
and away from it ($\delta \neq 0$).
The function    $ f_{\delta}(u)$
determines the full energy dependence of
the density of states and the localization
length (compare Eq.   (\protect\ref{gf})).
$I_0(x)$ and $I_1(x)$ are the zeroth and
first order modified
Bessel functions, respectively.
$\Gamma(\delta)$ is the gamma function.
$H_{\delta}^{(2)}(u)$ is a Hankel function of index $ \delta$.
$I_{iy}(x)$ is a modified Bessel function with imaginary index.
For small $x$, $\Gamma(x) \sim x^{-1}$,
$I_0(x) \sim 1 + x^2/4$, and $I_1(x) \sim x$.
  } \begin{tabular}{lcccc} Quantity & Symbol &\multicolumn{2}{c}{Commensurate}& Incommensurate \\
& & $\delta \neq 0$ & $\delta=0$ &\\
\tableline
\vtop{\baselineskip=10 pt \halign{#\hfil \cr
Density \cr of states \cr}}& $\langle \rho(E) \rangle/\rho_0  $
& $\displaystyle {\delta \over \Gamma(\delta)^2}
\left| { E \over  D  }\right| ^{2 \delta - 1}$
& $\displaystyle{ D  \over |E|
\left[\ln \left| { E \over D   }\right|
\right]^3 }$
& $\displaystyle{1 \over I_0(2\delta)^2}$\\
Specific heat & $\langle C(T) \rangle $ &
 $\displaystyle{ { \delta \over \Gamma(\delta)^2} \left({T \over D}
\right)^{2\delta} } $&
$\displaystyle{ 1 \over
\left[\ln \left( {T \over D }\right)
\right]^3 }$& $\displaystyle{T \over I_0(2\delta)^2}$\\
\vtop{\baselineskip=10 pt \halign{#\hfil \cr
Dynamical \cr critical exponent \cr}}& $z$ &
$\displaystyle{1 \over 2 \delta }$
 & $\infty$ &1 \\ \vtop{\baselineskip=10 pt \halign{#\hfil \cr
Localization \cr length \cr}}& $\lambda(E)$
&$\displaystyle{v_F \over \Delta }$& 
$\displaystyle{{v_F \over D}
\ln \left| {D \over E }\right| }$ &  
$\displaystyle{{4 v_F \over D}
\left[ 1 + {4\delta I_1(2\delta) \over I_0(2\delta) }\right]^{-1} }$ \\
\vtop{\baselineskip=10 pt \halign{#\hfil \cr
Transverse \cr Susceptibility \cr}}& $\langle \chi_{zz}(T)\rangle$
& $ T^{2\delta-1}$ & $\displaystyle{
1\over T \left[\ln \left( {T \over D}\right)
\right]^2 }$& $\displaystyle{1 \over I_0(2\delta)^2}$\\
& $ f_{\delta}(u)$
 & \multicolumn{2}{c}{
$\displaystyle {-u {\partial \over \partial u}\ln
(H_{\delta}^{(2)}(u)) }$} & 
$\displaystyle {\delta {\partial \over \partial \delta}
\ln (I_{-2iu}(2\delta))}$ \\
\end{tabular}
\label{table2}

\end{table}


\begin{references}

\bibitem[*]{email}electronic address: ross@newt.phys.unsw.edu.au

\bibitem{griffiths}
R.~B.~Griffiths, Phys. Rev. Lett. {\bf 23}, 17 (1969).

\bibitem{ry}
H. Rieger and   A. P. Young, cond-mat/9607005.

\bibitem{dsf}
D.~S.~Fisher, Phys. Rev. Lett. {\bf 69}, 534 (1992);
 Phys. Rev. B {\bf 51}, 6411 (1995).

\bibitem{ma}
C. Dasgupta and S.-k. Ma, Phys. Rev. B {\bf 22}, 1305 (1980).

\bibitem{yr}
A. P. Young and H. Rieger,   Phys. Rev. B {\bf 53}, 8486 (1996).

\bibitem{hyman}
R. A. Hyman {\it et al.}, Phys. Rev. Lett. {\bf 76}, 839 (1996).

\bibitem{dsf2}
D.~S.~Fisher, Phys. Rev. B {\bf
50}, 3799 (1994).

\bibitem{west}
E. Westerberg {\it et al.},
Phys. Rev. Lett. {\bf 75}, 4302 (1995).

\bibitem{senthil}
T. Senthil and S. N. Majumdar,
Phys. Rev. Lett. {\bf 76}, 3001 (1996).

\bibitem{bul}
L. N. Bulaevskii {\it et al.},
Zh. Eksp. Teor. Fiz. {\bf 62}, 725  (1972)
[Sov. Phys. JETP {\bf 35}, 384 (1972)].


\bibitem{nguyen}
T. N. Nguyen, P. A. Lee, and H.-C. zur Loye,
Science {\bf 271}, 489 (1996).

\bibitem{however}
However, it should be noted that in the fermion representation 
 results involving the magnetization  and its correlations
cannot be derived because the of the nonlocal relationship between the spin
and the fermion operators.

\bibitem{dennijs} M. den Nijs,
in {\it Phase transitions and critical phenomena},
edited  by C. Domb and J. L. Lebowitz
(Academic, London, 1988), vol. 12, p. 264.

\bibitem{pfeuty}
P.~Pfeuty, Ann. Phys. (NY) {\bf 27}, 79 (1970).

\bibitem{lsm}
E.~Lieb, T. Schultz and D.~Mattis, Ann. Phys. (NY) {\bf 16}, 407 (1961);
S.~Katsura, Phys. Rev. {\bf 127}, 1508 (1962).


\bibitem{dirac}
An easy way to derive (\protect\ref{hamel}) 
for the transverse Ising model in a random
transverse field is to take the continuum
limit of equation (39) in Ref. \protect\onlinecite{yr}.
For the disorder-free case see,
R.~Shankar, Acta Phys.   Pol. B {\bf 26}, 1835 (1995).

\bibitem{ov} A. A. Ovchinnikov and N. S. \'Erikhman,
Zh. Eksp. Teor. Fiz. {\bf 73}, 650  (1977)
[Sov. Phys. JETP {\bf 46}, 340 (1977)].

\bibitem{halp}
I. M. Lifshits, S. A. Gredeskul, and L. A. Pastur, {\it Introduction to
the Theory of Disordered Systems} (Wiley, New York, 1988), p.~109.

\bibitem{gor} L. P. Gor'kov and O. I. Dorokhov,
Fiz. Nizk. Temp.  {\bf 4}, 332 (1978)
[Sov. J. Low Temp. Phys. {\bf 4}, 160 (1978)].

\bibitem{golub} A. A, Golub and Y. M. Chumakov,  
Fiz. Nizk. Temp.  {\bf 5}, 900 (1979)
[Sov. J. Low Temp. Phys. {\bf 5}, 427 (1980)].

\bibitem{hayn}
R. Hayn and W. John, Z. Phys. B {\bf 67}, 169 (1987).

\bibitem{john}
H. J. Fischbeck and R. Hayn, Phys. Stat. Sol. (b) {\bf 158}, 565 (1990). 

\bibitem{mertsching}
J. Mertsching, Phys. Stat. Sol. (b) {\bf 174}, 129 (1992). 

\bibitem{mck}
For another application of this model,
R. H. McKenzie and J. W. Wilkins,
Phys. Rev. Lett. {\bf 69}, 1085 (1992).

\bibitem{bouch}
A mathematically equivalent model describes    
diffusion of a classical particle in
a random force field.
J. P. Bouchaud {\it et al.}, Ann. Phys. (NY) {\bf 201}, 285  (1990).

\bibitem{thou} D. J. Thouless, J. Phys. C {\bf 5}, 77 (1972).

\bibitem{smith}
E. R. Smith, J. Phys. C {\bf 3}, 1419 (1970).

\bibitem{scale}
The total number of states (per unit length) with energy less than
$E$, $N(E)\equiv \int_0^E \rho(E^\prime)dE^\prime$ scales with the inverse 
of any length scale $\ell$. By definition $E \sim \ell^z$.

\bibitem{mon}
C. Monthus {\it et al.},
Phys. Rev. A {\bf 54}, 231 (1996).             

\bibitem{klein} A. Klein
and J. F. Perez, Commun. Math. Phys. {\bf 128}.
99 (1990).

\bibitem{sm}
R.~Shankar and G.~Murphy, Phys. Rev. B {\bf 36}, 536 (1987).

\bibitem{altshuler}
B. L. Altshuler  and V. N. Prigodin,
Zh. Eksp. Teor. Fiz. {\bf 95}, 348 (1989)
[Sov. Phys. JETP {\bf 68}, 198 (1989)].

\end{references}
\end{document}